\documentclass[12pt]{article}
\usepackage{amsmath,graphicx}
\usepackage{hyperref}
\usepackage[utf8]{inputenc}
\usepackage[T1]{fontenc}
\usepackage{authblk}
\DeclareMathOperator\sen{sen}

\DeclareMathOperator\sign{sign}

\def\l({\left(}
\def\r){\right)}

\title{Collision frequency among non maneuverable objects}

\author[1]{Ricardo García-Pelayo\thanks{E-mail: r.garcia-pelayo@upm.es}}
\author[1]{E.K.V. Fernández de la Cigoña\thanks{E-mail: estanislaodekostka.vara@alumnos.upm.es}}
\author[1]{J. C. Velerda Escobar\thanks{%
E-mail: jcvelerda@indra.es.\\
\hspace*{1.9em}Currently at Indra Space, Avenida de Bruselas 35, Alcobendas 28108, Spain.}}
\author[2]{Shaker Bayajid Khan\thanks{E-mail: bayajid.khan@upm.es}}
\author[1]{S. Ocaña Losada\thanks{E-mail: s.ocanal@alumnos.upm.es}}

\affil[1]{ETS de Ingeniería Aeronáutica\\
Plaza del Cardenal Cisneros, 3\\
Universidad Politécnica de Madrid\\
Madrid 28040, Spain}
\affil[2]{Independent Researcher\\
81243 Munich, Germany}

\date{}

\begin{document}
\maketitle

\tableofcontents

\begin{abstract}

We compute the collision frequency among non maneuverable objects in May 2016, May 2021 and May 2026. The inputs for the computation of this frequency are the orbital parameters and sizes of the said objects. The computed frequencies are consistent with the frequency of the known collisions. They are inconclusive with respect to the question of whether we are approaching or not the Kessler cascading. A semi-empiric formula is found which gives the collision frequency as linear function of the square of the number of non maneuverable objects times the sum of the square of the average size and the variance of the size. We believe that computation of the collision frequency with better, future or simulated inputs will bring more conclusive results pertaining the Kessler cascading.

\end{abstract}

{\bf{

\centerline{}

}}
\vspace{0.5in}

{\centerline{\Large{\bf{Notation}}}}
\bigskip
$$
\begin{array}{cc}
  {\text{LEO}} & {\text{low Earth orbit}} \\
  J_2 & {\text{second-degree zonal harmonic coefficient}} \\
  \Omega & {\text{longitude of the ascending node}} \\
  \omega & {\text{argument of the perigee}} \\
  {\text{TLE}} & {\text{two line elements}} \\
  i & {\text{inclination}} \\
  {\text{pdf}} & {\text{probability density function}} \\
  q & {\text{perigee}} \\
  Q & {\text{apogeee}} \\
  \theta & {\text{true anomaly}} \\
  a & {\text{major semiaxis}} \\
  e & {\text{eccentricity}} \\
  \varphi & {\text{latitude}} \\
  p & {\text{semilatus rectum}} \\
  b & {\text{bearing}} \\
  \mu & G M_{Earth} \\
  h & {\text{angular momentum per unit mass}} \\
  GCAT & {\text{General Catalogue of Artificial Space Objects}} \\
\end{array}
$$

\section{Introduction}

In 1978 Kessler and Cour-Palais \cite{Kessler1978}, probably inspired by Kessler's previous work on the asteroid belt \cite{Kessler1971}, pointed out that repeated collisions among spatial debris and satellites lead to a chain reaction that would create a debris belt around the Earth that would forbid the LEO to satellites for many decades, perhaps as soon as companies perceive that putting a satellite in LEO is bad business \cite{Adilov2018}. This chain reaction has been called the ``Kessler cascading'' or the ``Kessler syndrome''.

In April of 1976 there were 3,866 bodies orbiting the Earth. A sample of 125 of them was chosen to compute, among other quantities, a crucial one: the frequency of collision \cite{Kessler1978}. Back then evasive maneuvers were not contemplated. In 1981 Kessler \cite{Kessler1981} computed frequencies of collisions between some Jupiter's moons using not statistical methods as in 1978, but the known orbital parameters and sizes of the said moons.

There has been since articles \cite{Sehnal1981,Hall2002,Liou2006b,Braun2020,Zhang2023} computing collision frequencies following the methods of references \cite{Kessler1978,Kessler1981} or similar ones. Some of these articles have also used the expected number of fragments produced by a collision to make forecasts about the Kessler cascading. None of these articles, however, have straightforwardly applied Kessler's methods of 1981 to the objects orbiting the Earth (instead of to the Jupiter Moons).

Whenever the next collision happens is, of course, a deterministic event. The computation of this instant has a forbidding cost. In the present article, we approximate this deterministic event by a random event, whose frequency can be computed at a reasonable cost.


We know that due to the oblateness of the Earth (technically, due to $J_2 \neq 0$), $\Omega$ and $\omega$ precess. Indeed, that is why they are not given in the two line elements (henceforth, TLE \cite{spacetrack_web}). We take them to be uniformly distributed and independent of each other. Under this assumption, each orbit is substituted by some sort of barrel limited by the altitudes corresponding to the perigee and the apogee and by the inclinations $\pm i$ (see Fig. \ref{fig:tonel}). The density inside each barrel is not uniform, but needs to be computed carefully. Once this density is known, the radii of each of the objects and their relative velocity are what we need to find the probability of collision. This is what was done by Kessler in 1981 for a few Jupiter moons and what we do here for tens of thousands of objects orbiting the Earth.

Suppose that the inclination, the perigee, and the apogee of the orbiting objects never changed. Suppose furthermore that when collisions happened, the objects would magically just go past each other without breaking, so that the current distribution of altitudes and inclinations would remain forever and the objects would just keep on colliding without consequences. The frequency of these events is what we find in the current article. The relation between the frequency and the probability that the next collision happens in a certain interval of time is discussed in subsection \ref{subsection:Frequencies of collision}.

We have computed the frequency of collision between all non maneuverable bodies orbiting the Earth in May of 2016, 2021 and 2026. We have, however, presented the method in far greater detail than Kessler in 1981 \cite{Kessler1981}. The data have been taken from the TLE of SATTRACK (US Space Force) \cite{spacetrack_web} and DISCOS (ESA) \cite{esa_discos_web}.

In sections 2-9 the calculations to be performed are explained, in section 10 the input data are discussed, in sections 11-14 the results are presented, discussed and successfully compared with the number of actual collisions which have taken place. A semi-empirical formula is presented for quick estimation of the collision frequency with about 10$\%$ error.

\section{Assumptions of the model}

Only the Earth's gravitational force is taken into account. Non gravitational forces (light pressure, friction with the atmosphere) and gravitational pull from other bodies are neglected. The Earth's mass distribution is expanded in spherical harmonics and only the $J_2$ term is kept.

It is remarkable that the model does not depend on the value of $J_2$. That is, as long as $J_2 \neq 0$, the Earth's oblateness will cause the Keplerian orbit to precede and this is enough to average over the orbital parameters $\Omega$ and $\omega$. Actually, the model has a weak dependence on $J_2$ via the relative velocity. But for small values of $J_2$, such as the Earth's, the contribution of the precession to the overall velocity can be neglected with respect to the Keplerian velocity.

So far the obvious physical assumptions. But there are two assumptions in the model to be presented which entail an underestimation of the frequencies of collision. They are discussed in the following two paragraphs.

We have restricted this study to the so called defunct objects, that is, non maneuverable objects. Thus the frequencies found in this article are lower bounds, but very close to reality. Indeed, while there have been cases in the past where maneuverable objects have collided (see the tables in section \ref{Comparison with observed frequencies}), it is remarkable that none of the satellites belonging to the constellation Starlink has collided. They have implemented an autonomous avoidance system; between December 2023 and May 2024 the constellation performed nearly 50,000 maneuvers in six months, equivalent to about 275 per day \cite{space_com}. We assume that this is how the future is going to look like: debris will be detected and evasive maneuvers will be performed successfully. This is not overly optimistic, given the improvements in surveillance \cite{ESA2025}.

We further assume that only objects above 10 cm diameter can cause catastrophic collisions, that is, one leading to fragmentation: ``The consequences of meteoroid and debris impacts on spacecraft can range from small surface pits due to micrometer-size impactors and clear-hole penetrations for millimeter-size objects, to mission-critical damage for projectiles larger than 1 cm. Any impact of a 10 cm catalogue object on a spacecraft or orbital stage will most likely entail a catastrophic disintegration of the target'' \cite{esa_space_safety}. However, objects with diameter under 10 cm may also cause catastrophic collisions. A more fundamental criterion for the catastrophic collision threshold is the NASA Standard Breakup Model, also called the 40 J/g criterion \cite{Johnson2001}. Thus an object smaller than 10 cm might break another one, if the other one is light enough. This criterion is again incomplete, because it does not take the relative speed, the material and geometry of the colliding bodies into account \cite{Francesconi2022}. Thus, as the previous one, this assumption underestimates the frequencies of collision.

The last two assumptions are, so to speak, mandatory. If collisions involving maneuverable objects were taken into account, the model would need to introduce a number of unknown parameters which would be quite difficult to guess. These parameters would quantify the different policies implemented by various operators when a conjunction warning is given. As for the 10 cm threshold, most objects which are smaller not tracked by radar and thus no TLE are provided for them.


\bigskip

\section{Probability density function of a moving point}

We are going to need the probability of finding a satellite in a given location. With this goal in mind we define the probability density function (henceforth, pdf) $\rho$ of a finding a moving point at $\vec r$ by
\begin{equation}\label{}
\rho(\vec r)\ \lim_{dr \to 0} B(\vec r,\ dr) = \lim_{T \to \infty} {\lim_{dr \to 0} t\big( B(\vec r,\ dr) \big) \over T},
\end{equation}
where $B(\vec r,\ dr)$ is the volume of the ball of center $\vec r$ and radius $dr$ and $t\big( B(\vec r,\ dr) \big)$ is the time spent in the ball of center $\vec r$ and radius $dr$ during an interval $T$. Of course the above limit does not always exist, but it exists in the instances in which we are going to use it. When the motion is periodic then the limit $T \to \infty$ does not need to be taken, the definition $\rho(\vec r)\ \lim_{dr \to 0} B(\vec r,\ dr) = {\lim_{dr \to 0} t( B(\vec r,\ dr) ) \over T}$ will do if $T$ is the period. Alternatively,

\begin{equation}\label{}
\rho(\vec r)\ \lim_{dr \to 0} B(\vec r,\ dr) = \lim_{T \to \infty} {\lim_{dr \to 0} \ell\big( B(\vec r,\ dr) \big) \over v(\vec r) T},
\end{equation}
where $v(\vec r)$ is the speed at $\vec r$ and $\ell\big( B(\vec r,\ dr) $ is the length of the trajectory caught by the ball of center $\vec r$ and radius $dr$.

When we are interested in a linear density $\lambda$ which depends only on one coordinate $x$ the above formula becomes

\begin{equation}\label{}
\lambda(x)\ \lim_{dx \to 0} dx = \lim_{T \to \infty} {\lim_{dx \to 0} x(dx) \over v(x) T}\ \Rightarrow\ \lambda(x) \propto {1 \over v(x)}
\end{equation}
and if the speed is known as a function of the position, then the normalization provides $\lambda(x)$:

\begin{equation}\label{dens1dim}
\lambda(x) = {1 \over v(x) \int dx'\ (1/v(x'))}.
\end{equation}

The general idea to find $\rho$ is the following.

Due to the $J_2$ precession both the longitude of the ascending node ($\Omega$) and the argument of the perigee ($\omega$) change. In the generic case $\dot \Omega$ and $\dot \omega$ are not rationally related and we may average over them independently.

\begin{figure}[ht]
  \begin{center}
 \includegraphics[width=0.45 \textwidth]{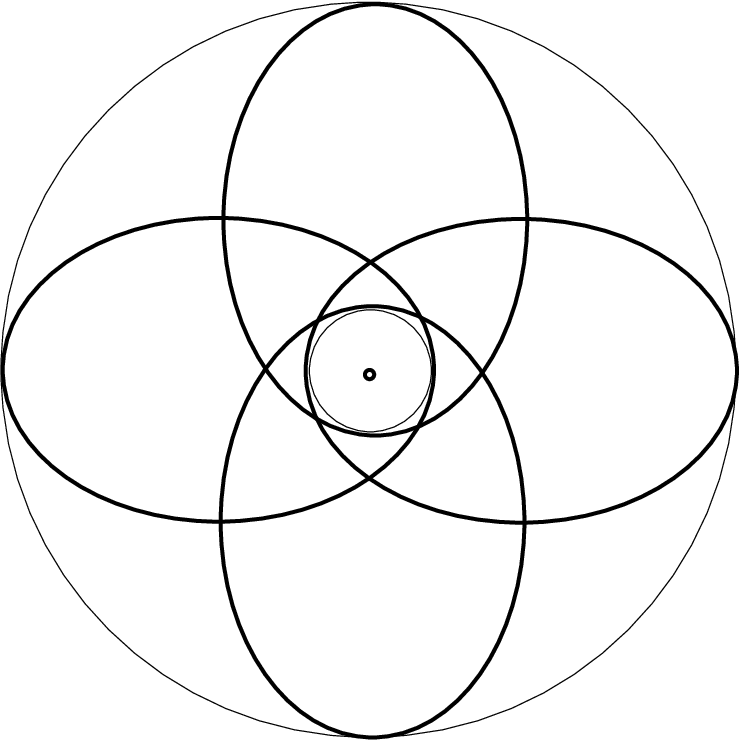} \caption{\label{fig:Prec.omega.1} Four orbits separated by $\pi/2$ angles from each other are shown.}
  \end{center}
\end{figure}

First we average over $\omega$. Then the Keplerian ellipse becomes a circular crown, with center at the occupied focus of the ellipse and radii $q$ and $Q$, the perigee and the apogee, respectively (see Fig. \ref{fig:Prec.omega.1}). There is a probability density, $\sigma(r)$, which has this crown as support.

%

This circular crown makes an angle $i$ with the Equatorial plane. Its revolution about the polar axis sweeps a volume which is the intersection between a spherical shell of radii $q$ and $Q$ and the outside volume a cone whose generating lines make angles $\pm i$ with the Equatorial plane. This intersection is similar to a barrel (see Fig. \ref{fig:tonel}).

\section{Distribution of the radius in an elliptic orbit}

\begin{figure}[ht]
  \begin{center}
  \includegraphics[width=0.65 \textwidth]{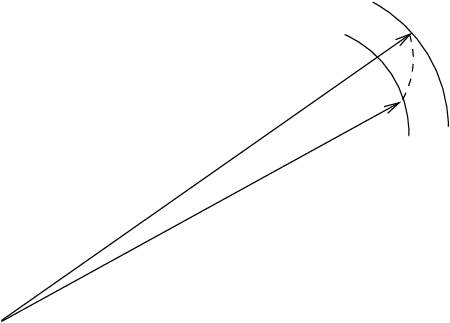} \caption{\small{One can see that the radial density is $\sigma(r) \propto {dt \over dr} = {1 \over \dot r}$. Note that this result is general. In particular, it does not require the orbit to be elliptic or hyperbolic.
  }}
  \end{center}
\end{figure}


Let $\lambda_K(r)$ be the radial pdf of a Keplerian elliptical orbit. According to formula \eqref{dens1dim} the essential task to find $\lambda_K(r)$ is to find the radial speed of a Keplerian elliptical orbit as function of the radius.

In this section we use the polar form of the elliptical orbit with origin in the occupied focus

\begin{equation}\label{FormaPolar2}
r(\theta)=\frac{a (1-e^2)}{1 + e\cos\theta},
\end{equation}
where the $\theta$ is the true anomaly, $a$ is the major semiaxis and $e$ is the eccentricity. Then

\begin{equation}\label{}
\dot r(\theta) = \frac{a (1-e^2)\ e \sin \theta}{(1 + e\cos\theta)^2}\ \dot \theta.
\end{equation}
But, due to the constancy of the areal velocity, $\dot \theta \propto {1 \over r^2}$. Thus, using the two previous equations,

\begin{equation}\label{}
\dot r(\theta) \propto \frac{\sin \theta}{r^2 (1 + e \cos\theta)^2} \propto \sin \theta,
\end{equation}
since the denominator of the second expression, as seen in the polar form \eqref{FormaPolar2}, is constant. From the polar form $\sin \theta$ can be solved:

$$
\sin \theta =
\sqrt{1 - \cos^2 \theta} = \sqrt{1 - {1 \over e^2} + 2 {a (1 - e^2) \over r e^2} - {a^2 (1 - e^2)^2 \over r^2 e^2}} =
$$

$$
{\sqrt{1 - e^2} \over r e^2} \sqrt{-r^2 + 2 a r - a^2 (1 - e^2)} =
{\sqrt{1 - e^2} \over r e^2} \sqrt{\big( a (1+e) - r \big) \big( r - a (1-e) \big)} =
$$

\begin{equation}\label{}
{\sqrt{1 - e^2} \over r e^2} \sqrt{\big( Q - r \big) \big( r - q \big)}.
\end{equation}


The normalization condition $\int dr\ \rho(r) = 1$ yields that the radial density is:


\begin{equation}
\lambda_K(r) = {r \over \pi a  \sqrt{(Q - r) (r - q)}} = {2 r \over \pi (q+Q)  \sqrt{(Q - r) (r - q)}},
\end{equation}
in accordance with \cite{Kessler1981}.

%

\section{Density of a preceding orbit}


The pdf $\lambda_K(r)$ for $r \in [q, Q]$ is known. The normalization condition $\int_q^Q dr\ \lambda_K(r) = 1$ of $\lambda_K(r)$ can be turned into a normalization condition for the surface probability density $\sigma(r)$ on the circular crown

\begin{equation}\label{}
\int_q^Q dr\ \lambda_K(r) = \int_q^Q dr\ r \int_0^{2 \pi} d\nu\ {\lambda_K(r) \over 2 \pi r} = 1\ \Rightarrow\ \sigma(r) = {\lambda_K(r) \over 2 \pi r}.
\end{equation}

Think of the circular crown of inclination $i$ whose intersection with the Equatorial plane is the $x$ axis. You can think of it as a circular crown which was originally in the Equatorial plane and was turned around the $x$ axis by an angle $i$ (see Fig. \ref{fig:bearing}). The image of any Equatorial point of coordinates $r (\cos \nu, \sin\nu)$ under this rotation is

\begin{equation}
r
\left(
  \begin{array}{ccc}
    1 & 0 & 0 \\
    0 & \cos i & -\sin i \\
    0 & \sin i & \cos i \\
  \end{array}
\right)
\left(
  \begin{array}{c}
    \cos \nu \\
    \sin \nu \\
    0 \\
  \end{array}
\right)
=
r
\left(
  \begin{array}{c}
    \cos \nu \\
    \cos i \sin \nu \\
    \sin i \sin \nu \\
  \end{array}
\right).
\end{equation}

The $z$ component of the position vector of a satellite is $r \sin \varphi$. Therefore, the relation between the angular variable $\nu = \omega + \theta$, the inclination $i$ and the latitude $\varphi$ is
\begin{equation}\label{SineRule}
\sin \varphi = \sin i \sin \nu \Rightarrow \sin \nu = \frac{\sin \varphi}{\sin i}.
\end{equation}

Taking derivatives on both sides:
\begin{equation}\label{nufi}
\cos \nu\ d \nu= {\cos \varphi \over \sin i}\ d\varphi \Rightarrow  {d \nu \over d \varphi} = {\cos \varphi \over \sin i \cos \nu} =
{\cos \varphi \over \sin i\ \sqrt{1 - {\sin^2 \varphi \over \sin^2 i}}} = {\cos \varphi \over \sqrt{\sin^2 i - \sin^2 \varphi}}.
\end{equation}
Relation \eqref{nufi} between $d \nu$ and $d \varphi$ is valid at each point. However, for a given latitude $\varphi$ there are two different angles $\nu$, therefore when the change of variable from $\nu$ to $\varphi$ is made, the correct substitution is $d\nu \rightarrow 2 {\cos \varphi d\varphi \over \sqrt{\sin^2 i - \sin^2 \varphi}}$, because one infinitesimal angle $d\varphi$ corresponds to two infinitesimal angles $d\nu$. From the normalization condition for the surface pdf $\sigma(r)$ we may obtain the wanted $\rho(r, \varphi)$:

$$
1 = \int_q^Q dr\ r \int_0^{2 \pi} d\nu\ {\lambda_K(r) \over 2 \pi r} = 2 \pi \int_q^Q dr\ \int_{-i}^{+i} d\varphi\ r^2 \cos \varphi {2 \over \sqrt{\sin^2 i - \sin^2 \varphi}} {\lambda_K(r) \over (2 \pi)^2 r^2} =
$$

\begin{equation}\label{}
\int_0^{2 \pi} d{\rm{longitude}}\ \int_{-i}^{+i} d\varphi\ \int_q^Q dr\ {r^2 \cos \varphi \over \pi^3 r (q+Q) \sqrt{(Q - r) (r - q) (\sin^2 i - \sin^2 \varphi)}},
\end{equation}
therefore

\begin{equation}\label{eq:density}
\rho(r, \varphi) = {1 \over \pi^3 r (q+Q) \sqrt{(Q - r) (r - q) (\sin^2 i - \sin^2 \varphi)}}
\end{equation}
follows, which is formula (21) of \cite{Kessler1981}.

\begin{figure}[ht]
  \begin{center}
 \includegraphics[width=0.55 \textwidth]{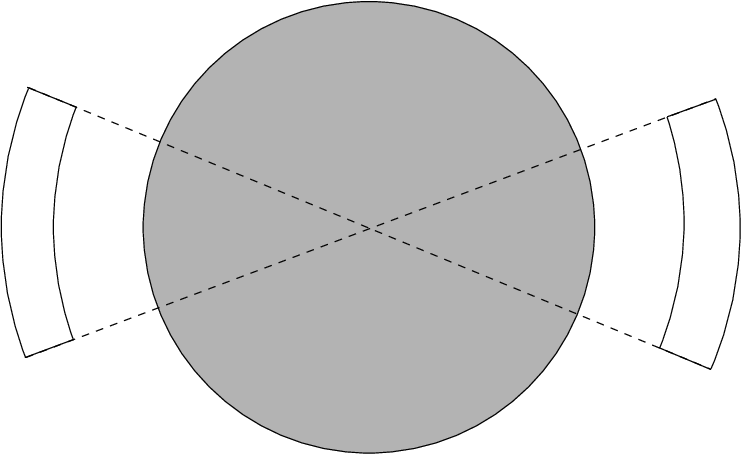} \caption{\label{fig:tonel} Section of the barrel. The Earth is shown in grey. The dotted lines make angles $\pm i$ with the Equatorial plane.}
  \end{center}
\end{figure}

%
%

\section{Collision frequency}

Consider a mesoscopic volume in the region where the barrels of bodies 1 and 2 overlap. The volume is mesoscopic in the sense that it is much larger than the bodies involved but much smaller than the dimensions of the barrels. We consider both bodies to be spherical of radii $R_1$ and $R_2$. Then they collide iff the distance between their centers is smaller than

\begin{equation}\label{}
R \equiv R_1 + R_2.
\end{equation}
If $R_1$ becomes $R$ and $R_2$ becomes 0 (that is, a point particle) the collision events will be the same. We substitute then the original radii by $R$ and 0 in order to think more clearly.

\begin{figure}[ht]
\begin{center}
\includegraphics[width=0.37 \textwidth]{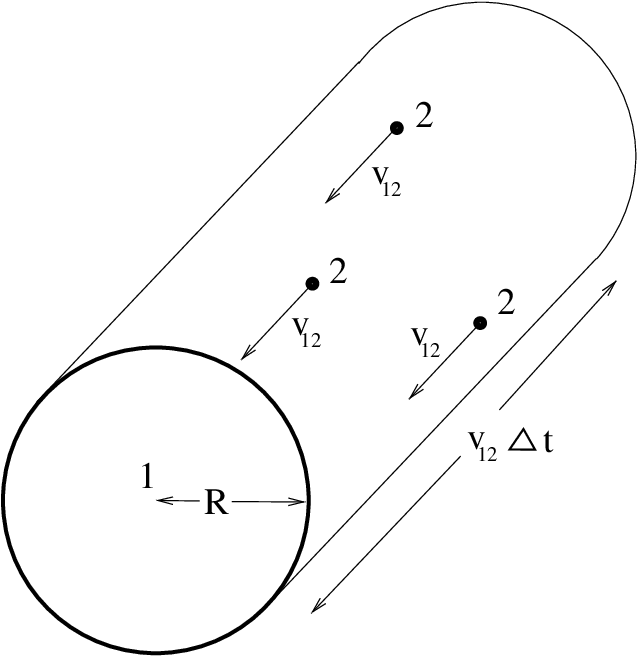}
\caption{\label{fig:AE} Cylinder containing the particles 2 which will hit the sphere 1 during $\Delta t$.
}
\end{center}
\end{figure}

The number of particles 2 incident upon a sphere of radius $R$ during a time $\Delta t$ is $\pi R^2 \rho_2 v_{12} \Delta t$, where $v_{12}$ is the relative speed of the particles. The probability that there is a sphere of type 1 in the volume $d^3r$ is $\rho_1(\vec r)\ d^3r$. Thus the collision frequency between bodies 1 and 2 is

\begin{equation}\label{}
{1 \over \tau_{12}} = \pi R^2 \int d^3r\ \rho_1(\vec r) \rho_2(\vec r) {v_{12+}(\vec r) + v_{12-}(\vec r) \over 2},
\end{equation}
where 
$v_{12\pm}(\vec r)$ stands for the two possible relative speeds of bodies 1 and 2 at position $\vec r$, which we shall find in section \ref{Relative speed}. While  ${v_{12+}(\vec r) + v_{12-}(\vec r) \over 2}$ is, in fact, the average of both speeds, one ought to think about this expression as the weighed probability (by the weight 1/2) of each case.

We have not specified the region over which the integration takes place. This region is the intersection of the barrels corresponding to bodies 1 and 2. In spherical coordinates it is clear that there is no dependence on the longitude - because of the precession - and that the limits of integration over $r$ are max$(q_1, q_2)$ and min$(Q_1,Q_2)$, provided max$(q_1, q_2) < $ min$(Q_1,Q_2)$. To determine the limits of integration over the latitude we need to proceed a little more carefully. The first thing that comes to mind is that the limits are $\pm$min$(i_1,i_2)$. This is right if both orbits are prograde (that is, $i \leq \pi/2$). One or both may, however, be retrograde (that is, $i > \pi/2$). To cover this case we define the function $f(i)$

\begin{equation}\label{eq:I}
I(i) =
\left\{
\begin{array}{ccc}
  i & i < \pi/2 \\
  \pi - i & i \geq \pi/2
\end{array}
\right.
\end{equation}
Then the integration domain for $\varphi$ is [$-$min$(I_1, I_2)$, +min$(I_1, I_2)]$ $\big({\rm{where}}\ (I_1,I_2) \equiv (I(i_1), I(i_2)) \big)$ and the expression for the collision frequency between bodies 1 and 2 becomes

$$
{1 \over \tau_{12}} = \pi R^2 2 \pi \int_{\text{max}(q_1,q_2)}^{\text{min}(Q_1,Q_2)} dr\ r^2\ \int_{-\text{min}(I_1,I_2)}^{+\text{min}(I_1,I_2)} d\varphi\ \cos\varphi\ \rho_1(r, \varphi) {v_{12+}(r, \varphi) + v_{12-}(r, \varphi) \over 2} \rho_2(r, \varphi) =
$$

\begin{equation}\label{eq:CollRate}
2 \pi^2 R^2 \int_{\text{max}(q_1,q_2)}^{\text{min}(Q_1,Q_2)} dr\ r^2\ \int_{0}^{\text{min}(I_1,I_2)} d\varphi\ \cos\varphi\ \rho_1(r, \varphi) \big[ v_{12+}(r, \varphi) + v_{12-}(r, \varphi) \big] \rho_2(r, \varphi),
\end{equation}
when max$(q_1, q_2) < $min$(Q_1,Q_2)$ and it is 0 otherwise.

The last form of ${1 \over \tau_{12}}$ follows from the fact that for each collision over the Northern Hemisphere at coordinates $(\lambda, \varphi)$ there is a collision with the same relative speed in the Southern Hemisphere at coordinates $(\lambda + \pi, -\varphi)$, as shown in Fig. \ref{fig:bearing}. Alternatively, one can see from the analytic form \eqref{eq:vrel} of $\big[ v_{12+}(r, \varphi) + v_{12-}(r, \varphi) \big]$, that it is invariant under $+\varphi \leftrightarrow -\varphi$.

In order to find the relative speed, the cosine of the angle between orbital planes will be needed.


\section{Angle between orbital planes} \label{Angle between orbital planes}

\begin{figure}[ht]
  \begin{center}
  \includegraphics[width=0.6 \textwidth]{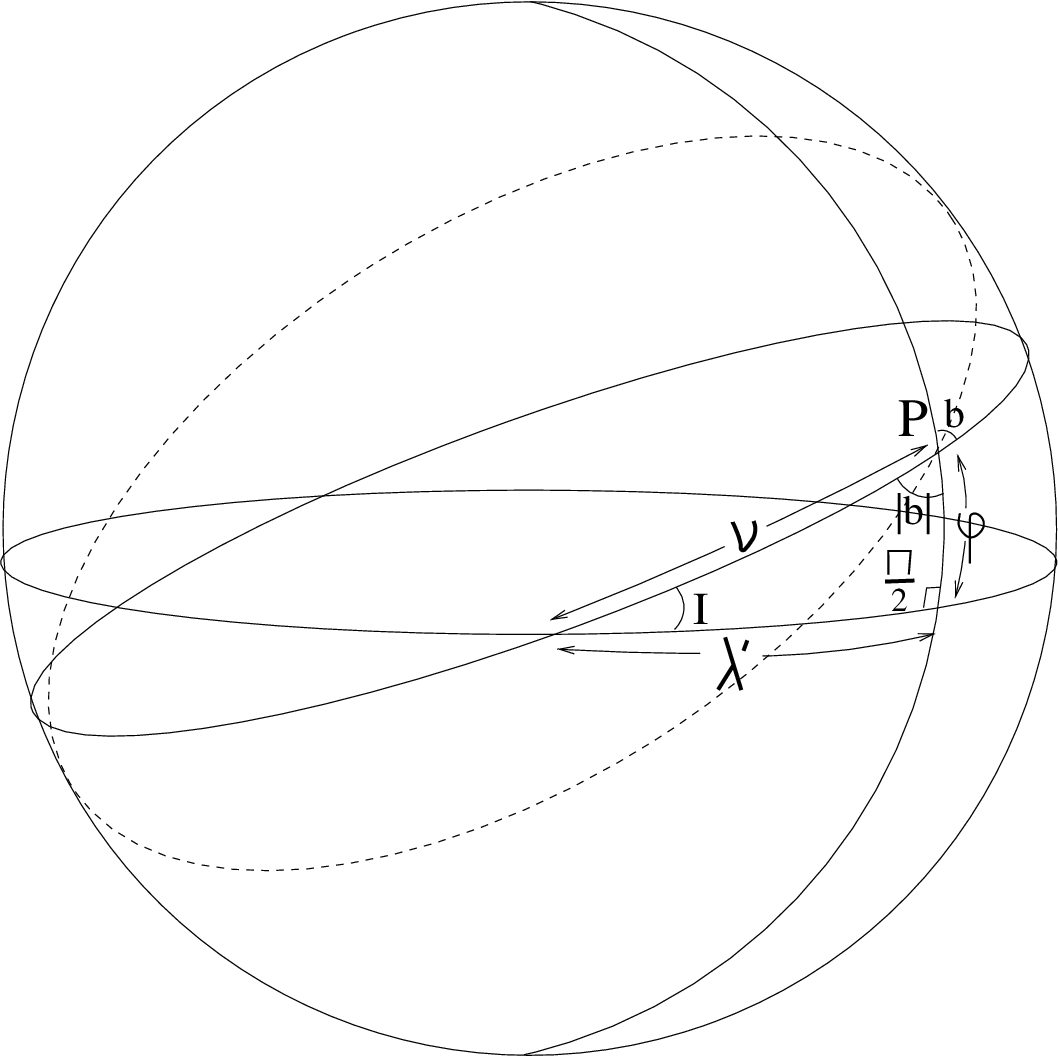} \caption{\label{fig:bearing} The circular projection on a sphere of an elliptical orbit is shown. This projection has another point (not shown) on the back side of the Earth, which has the same latitude $\varphi$. Note that $\nu = \omega + \theta$. At the point shown, for which $\lambda'$ and $\nu$ are in $[0,\pi/2]$, the latitude is increasing and $\sin b >0$. For the point not shown $\lambda'$ and $\nu$ are in $[\pi/2, \pi]$, the latitude is decreasing and $\sin b < 0$. The dotted circle is the projection on the sphere of a different elliptical orbit. The angle between their velocity vectors is the difference of their bearings. In the prograde orbit shown, the angles of the triangle which are not the straight angle are $i$ and $b$. For retrograde orbits, however, the corresponding angles are $I \equiv \pi - i$ and $|b|$.}
  \end{center}
\end{figure}

Let $b$ denote the {\it{bearing}}, that is, the angle made by the horizontal projection of the velocity at a given point with the North direction of the meridian passing through that point, as defined in navigation \cite{Keay1989} (see Fig. \ref{fig:bearing}). The range of $b$ is $[-\pi, +\pi]$. When $b<0$ the motion is westward and when $b>0$ the motion is eastward, that is, for prograde orbits ($i \in [0, \pi/2]$) the bearing is positive and for retrograde orbits ($i \in [\pi/2, \pi]$) the bearing is negative. When $|b|<\pi/2$ the motion is to the North and when when $|b|>\pi/2$ the motion is to the South. Clearly, ${\pi \over 2} - b = i$ when $\varphi = 0$ (in the Equator), $b = {\pi \over 2}$ when $\varphi = \pm i$ (at the furthermost North and South points) for prograde orbits and $b = - {\pi \over 2}$ when $\varphi = \pm i$ for retrograde orbits. We are interested in the bearing because the angle made by two orbital planes is the angle made by their bearings at their points of intersection.

For any point $P$ we define the arc of angle $\lambda'$ as the arc of the Equator which goes in the positive direction (to the East) from the ascending node to the meridian which passes through the point $P$; $\lambda' \in [0, 2 \pi]$. We define the arc of angle $\nu$ as the arc of circle which goes in the positive direction (to the East) from the ascending node to the point $P$; $\nu \in [0, 2 \pi]$. Notice that the signs of $\sin \lambda'$ and $\sin \nu$ are always the same. When $\Omega = 0$, $\lambda'$ is the longitude of $P$ and when $\omega = 0$, $\nu$ is the true anomaly.

Any point $P$ makes a right triangle whose other vertices are the ascending node and the point on the Equator with the same longitude as $P$ (see Fig. \ref{fig:bearing}). The angles corresponding to these vertices are $\pi/2$, $I$ (its definition is \eqref{eq:I}) and $|b|$, respectively. The lengths of the arcs opposite to each vertex are (we take the radius of the sphere to be 1) $\nu$ (or $\nu-\pi$ when $\nu > \pi$) (the hypothenuse), $|\varphi|$ and $\lambda'$ (or $\lambda'-\pi$ when $\lambda' > \pi$) (the legs), respectively.



Our goal is to find the cosine $\cos(b_2 - b_1)$ of the angle between two orbital planes at a given point $(r, \varphi)$. To this end we are going to find the function $b(i, \varphi)$ first.

This relation is provided by the cosine law for angles of spherical trigonometry (formula (14) in p. 1700 of \cite{Weisstein1999}). This law reads:

\begin{equation}\label{}
\cos A = -\cos B \cos C + \sin B \sin C \cos a,
\end{equation}
where the usual convention of spherical trigonometry is used: capital letters for the angles and the same small case letter for the opposing arc. We set $C = \pi/2$, for the right triangle case. Then the formula simplifies to

\begin{equation}\label{}
\cos A = \sin B \cos a,
\end{equation}
(formula (6) in p. 37 of \cite{Todhunter1886}) which implies, for the right triangle described above,

%
%

\begin{equation}\label{eq:rumbo}
\sin b = {\cos i \over \cos \varphi},
\end{equation}
This formula holds for prograde and retrograde orbits. Indeed, when the orbit is retrograde the formula is $\sin |b| = {\cos I \over \cos \varphi}$, but in the retrograde case $\sin |b| = - \sin b$ and $\cos I = \cos i$, thus the formula holds.
%
It follows from formula \eqref{eq:rumbo} that the cosine of the angle of the horizontal components of two velocities between two orbital planes is

$$
\cos (b_2 - b_1) = \cos b_2 \cos b_1 + \sin b_2 \sin b_1 =
$$

\begin{equation}\label{eq:cosb1b2}
{1 \over \cos^2 \varphi} \left ( \pm \sqrt{\cos^2 \varphi - \cos^2 i_2} \sqrt{\cos^2 \varphi - \cos^2 i_1} + \cos i_2 \cos i_1 \right),
\end{equation}
where the $\pm$ sign will be discussed later (this sign is + if both bodies are North or South bound, and $-$ otherwise).

%

\section{Orbits through a point of given latitude} \label{Orbits through a point of given latitude}

We use the notation $\mu \equiv G M_{Earth}$ and $h$ for the angular momentum per unit mass of the body due to its orbital motion.

In a Keplerian orbit (except for the circular one) there are two true anomalies for a given $r$. Indeed, in the elliptic orbit case,

\begin{equation}\label{eq:theta}
r = { h^2/\mu \over 1 + e \cos \theta} \Rightarrow \cos \theta = {1 \over e} \left( {h^2 \over \mu r} - 1 \right).
\end{equation}
There are two solutions for $\theta$, symmetric with respect to the major axis of the ellipse and which have sines of opposite sign. For a prograde orbit, when $\theta \in [0, \pi] \Leftrightarrow \sin \theta > 0$, the body travels from the perigee to the apogee, and therefore it gains altitude. When $\theta \in [\pi, 2 \pi] \Leftrightarrow \sin \theta < 0$, the opposite is true.

We now turn our attention to the sign of $\cos b$. It changes when the body reaches its minimum or maximum latitude. A given orbit has two points of the same latitude. In Fig. \ref{fig:bearing} one of them is shown, while the other one is on the back side of the Earth.
Suppose first that, as in Fig. \ref{fig:bearing}, the latitude $\varphi$ is positive. Then the first position after the ascending node at which the body reaches $\varphi$ is one of increasing latitude, that is, $\cos b > 0$, while the second is of decreasing latitude, that is, $\cos b < 0$. The reverse is true if the latitude is negative. These statements are true both for prograde and retrograde orbits.

Since both positions which have a given positive $\varphi$ are within the same hemisphere, it follows that when both are joined by the shortest geodesic with the direction of the motion, the first position after the ascending node is of increasing $\varphi$ while the second is of decreasing $\varphi$. The reverse is true if the latitude is negative.

\begin{figure}[ht]
\begin{center}
\includegraphics[width=0.43 \textwidth]{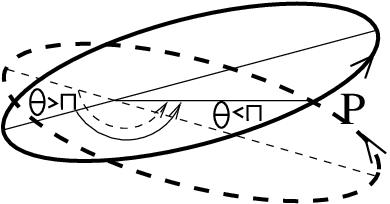}\hspace{0.07 \textwidth}\includegraphics[width=0.5 \textwidth]{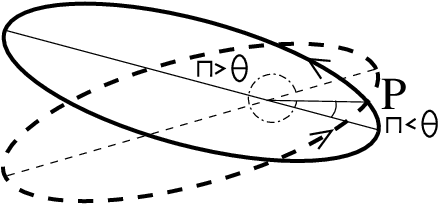}
\caption{\label{fig:Sols} Four prograde orbits, as seen from the Northern Pole Star, which go through the point $P$ of positive latitude (for typographical reasons the occupied focii have been drawn closer to the center of their ellipses than they actually are). We denote their {\it{semilatus rectum}} by $p$. When $r>p$ (left figure) and $P$ is the first position of a given latitude $\varphi$ met after the ascending node, then the altitude and latitude increase (continuous orbit, $\sin \theta > 0,\ \cos b >0$). When $r>p$ (left figure) and $P$ is the second position of a given latitude $\varphi$ met after the ascending node, it is of decreasing altitude and latitude (dashed orbit, $\sin \theta < 0,\ \cos b <0$). When $r<p$ (right figure) and $P$ is the first position of given latitude $\varphi$ met after the ascending node, then it is of decreasing altitude and increasing latitude (dashed orbit, $\sin \theta < 0,\ \cos b >0$). When $r<p$ (left figure) and $P$ is the second position of a given latitude $\varphi$ met after the ascending node, it is of increasing altitude and decreasing latitude (continuous orbit, $\sin \theta > 0,\ \cos b < 0$). The same rules are obtained for the retrograde orbits. When the point $P$ has negative latitude, opposite rules are obtained.}
\end{center}
\end{figure}


The relations between the sign of $\sin \theta$ and $\cos b$ are explained in Fig. \ref{fig:Sols}. The rules extracted from it are in the following tables:
\medskip

\hspace{0.1 \textwidth}
\begin{tabular}{|c|c|}
\hline
\multicolumn{2} { | c | }{$\varphi > 0$, N. Hemisphere}\\
\hline
  \hline
  $r < p$ & $r > p$ \\
  \hline
   $r \uparrow\ \varphi \downarrow$ & $r \uparrow\ \varphi \uparrow$ \\
  \hline
   $r \downarrow\ \varphi \uparrow$ & $r \downarrow\ \varphi \downarrow$ \\
  \hline
\end{tabular}
\hspace{0.14 \textwidth}
\begin{tabular}{|c|c|}
\hline
\multicolumn{2} { | c | }{$\varphi < 0$, S. Hemisphere}\\
\hline
  \hline
  $r < p$ & $r > p$ \\
  \hline
   $r \uparrow\ \varphi \uparrow$ & $r \uparrow\ \varphi \downarrow$ \\
  \hline
   $r \downarrow\ \varphi \downarrow$ & $r \downarrow\ \varphi \uparrow$ \\
  \hline
\end{tabular}
\medskip

Since $r \uparrow\ \Leftrightarrow \sin \theta > 0$ and $\varphi \uparrow\ \Leftrightarrow \cos b > 0$, it follows from the tables that the relative sign of $\sin \theta$ and $\cos b$ is

$$
\sign (\varphi) \sign(r-p).
$$
Let there be two orbits, of indices 1 and 2, which meet at a point of radius and latitude $r$ and $\varphi$ (the longitude is irrelevant). Then the relative sign of $\sin \theta_1 \sin \theta_2$ and $\cos b_1 \cos b_2$, to be used later, is

\begin{equation}\label{eq:s12}
s_{12}(r) \equiv \sign (\varphi) \sign(r-p_1) \sign (\varphi) \sign(r-p_2) = \sign(r-p_1) \sign(r-p_2).
\end{equation}

\section{Relative speed} \label{Relative speed}


To find the magnitude of the relative velocity, the $z$ axis is chosen as the radial direction at a point $P$, that is, the upward vertical direction. First, the $z$ component of the velocity difference will be found. Next, applying the cosine theorem will allow us to find the magnitude of the difference between the horizontal components of $\vec v_1$ and $\vec v_2$ perpendicular to $z$ (that is, the components along $\vec u_{\theta 1}$ and $\vec u_{\theta 2}$, respectively). Finally, by applying the Pythagorean theorem, we will arrive at the expression for the magnitude of the relative velocity.

\begin{figure}[ht]
\begin{center}
\includegraphics[width=0.5 \textwidth]{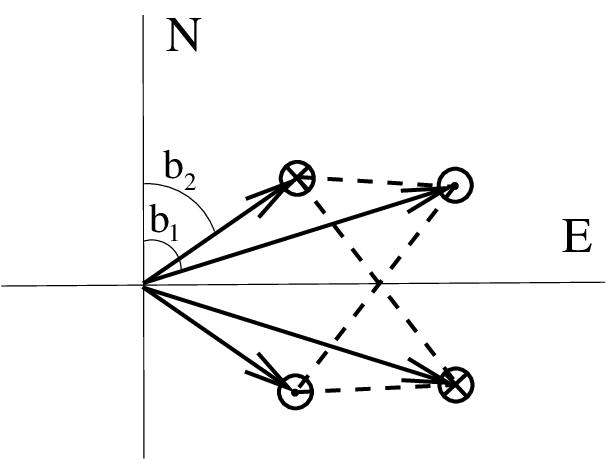}
\caption{\label{fig:RelSpeed} For a given latitude $\varphi$ and radius $r$ there are two possible velocities for each satellite. They are drawn with a Cartesian reference frame whose axes are the parallel, the meridian and the vertical. If the depicted velocities are of an object in prograde orbit in the Northern Hemisphere, then the two upper velocities are at a point of increasing latitude, and the two lower velocities are at a point of decreasing latitude.
It is shown in the figure that the four differences of velocities yield just two different speeds.}
\end{center}
\end{figure}

According to Appendix I, the velocity of the satellite in the $i$-th elliptical orbit is given by
\begin{equation}\label{eq:VelKepleriana}
\vec v_i = {\mu \over h_i} \left( e_i \vec u_{pi} + \vec u_{\theta i} \right),
\end{equation}
where $e_i$ is the eccentricity of the $i$-th orbit. The vector $\vec u_{pi}$ can be expressed as
\begin{equation}\label{}
\vec u_{pi} = {\sin \theta_i \vec k + \cos \theta_i \vec u_{\theta i}},
\end{equation}
where $\vec k = \vec u_r$ is the upward vertical unit vector.

Therefore the vertical components of the velocities are
\begin{equation}\label{}
{\mu e_i  \over h_i} \sin \theta_i \vec k.
\end{equation}
Note that the vector $\vec k$ is the same for both orbits. The difference between the two vertical components is
\begin{equation}\label{}
\mu  \left( {e_1 \over h_1} \sin \theta_1 - {e_2  \over h_2} \sin \theta_2 \right) \vec k,
\end{equation}
whose modulus squared is
\begin{equation}\label{}
\mu^2  \left( {e_1^2  \over h_1^2} \sin^2 \theta_1 + {e_2^2  \over h_2^2} \sin^2 \theta_2 - 2 {e_1  \over h_1} \sin \theta_1 {e_2  \over h_2} \sin \theta_2 \right).
\end{equation}

On the other hand, the horizontal components are
\begin{equation}\label{}
{\mu \over h_i} \left( 1 + e_i \cos \theta_i \right) \vec u_{\theta i}.
\end{equation}

The difference between these two horizontal components is
\begin{equation}\label{}
\mu \left( {1 + e_2 \cos \theta_2 \over h_2} \vec u_{\theta 2} - {1 + e_1 \cos \theta_1 \over h_1} \vec u_{\theta 1} \right),
\end{equation}
whose squared modulus is
\begin{multline}\label{}
    \mu^2 \left( {e_2^2  \over h_2^2} \cos^2 \theta_2 + {e_1^2  \over h_1^2} \cos^2 \theta_1 + {1 \over h_2^2} + {1 \over h_1^2} + 2 {e_2 \cos \theta_2 \over h_2^2} + 2 {e_1 \cos \theta_1 \over h_1^2} \right. \\
    \left. - 2 {1 + e_2 \cos \theta_2 \over h_2} {1 + e_1 \cos \theta_1 \over h_1} \cos (b_2 - b_1) \right),
\end{multline}
where $\cos (b_2 - b_1) = \vec u_{\theta 1} \cdot \vec u_{\theta 2}$ is the angle made by the two orbital planes, obtained in section \ref{Angle between orbital planes}.

The application of Pythagoras' theorem yields the relative speed

\begin{multline}\label{eq:vh12}
    v_{12} =
    \mu \left( {1 + e_1^2 \over h_1^2} + {1 + e_2^2  \over h_2^2} - 2 {e_1  \over h_1} \sin \theta_1 {e_2 \over h_2} \sin \theta_2 + 2 {e_2 \cos \theta_2 \over h_2^2} + 2 {e_1 \cos \theta_1 \over h_1^2} \right. \\
    \left. - 2 {1 + e_2 \cos \theta_2 \over h_2} {1 + e_1 \cos \theta_1 \over h_1} \cos (b_2 - b_1) \right)^{1 \over 2}.
\end{multline}

Using $\sin \theta = \pm \sqrt{1 - \cos^2 \theta}$ and formulae \eqref{eq:cosb1b2}, \eqref{eq:theta} and  \eqref{eq:s12}, one obtains, after a little algebra,


%
%

\begin{multline}\label{eq:vrel_notfinal}
    v_{12\pm} =
    \mu \Bigg( {4 \over \mu r} - {1 - e_1^2 \over h_1^2} - {1 - e_2^2 \over h_2^2} \pm {2 \over h_1 h_2} \sqrt{ e_1^2 - \Big( {h_1^2 - \mu r \over \mu r} \Big)^2 } \sqrt{ e_2^2 - \Big( {h_2^2 - \mu r \over \mu r} \Big)^2 } \\
    - 2 {h_1 h_2 \over \mu^2 r^2} {1 \over \cos^2 \varphi} \left ( \pm s_{12}(r) \sqrt{\cos^2 \varphi - \cos^2 i_2} \sqrt{\cos^2 \varphi - \cos^2 i_1} + \cos i_2 \cos i_1 \right) \Bigg)^{1 \over 2}.
\end{multline}
The first $\pm$ sign of the above expression corresponds to the two true anomalies which are solutions of equation \eqref{eq:theta}. The two terms which are products of square roots correspond to the products $\sin \theta_1 \sin \theta_2$ and $\cos b_1 \cos b_2$, respectively. Their relative sign is given by formula \eqref{eq:s12}. Thus the sign of the second product of square roots is not independent of the first, but related by the function $s_{12}$ (definition \eqref{eq:s12}). Therefore there are not $2 \times 2 = 4$ possible sign combinations, but just the two shown, that is, $+$ in both instances or $-$ in both instances. When one or both orbits are circular the above expression simplifies. Since these cases are not generic they are discussed in Appendix II.

%
%

Expression \eqref{eq:vrel_notfinal} is not yet ready for use because it is not given in terms of the parameters provided by the TLE. To this end, we transform the expression obtained for the relative speed $v_{12\pm}$ so that it depends only on the distance to the occupied focus, $r$, and the latitude at which the bodies are located, $\varphi$; as well as on the geometrical parameters of the orbits provided by the TLE: the distances to the apogee and perigee ($Q$ and $q$, respectively) and the inclination ($i$).

The expressions that relate the mentioned parameters with those that appear in formula \eqref{eq:vrel_notfinal} are:
\begin{equation}\label{MOaGeom1}
p = {h^2 \over \mu} = {2 q Q \over q + Q}
\end{equation}
and
\begin{equation}\label{MOaGeom2}
e = {Q - q \over q + Q}.
\end{equation}
Before making the above substitutions in expression \eqref{eq:vrel_notfinal}, some preliminary simplifications will be made. First, the fourth term of \eqref{eq:vrel_notfinal} is developed:

\begin{gather*}
    {2 \over h_1 h_2} \sqrt{ e_1^2 - \Big( {h_1^2 - \mu r \over \mu r} \Big)^2 } \sqrt{ e_2^2 - \Big( {h_2^2 - \mu r \over \mu r} \Big)^2 } = \\
    = {2 \over h_1 h_2} \sqrt{e_1^2 - {h_1^4 \over \mu^2 r^2} + {2 h_1^2 \over \mu r} -1} \sqrt{e_2^2 - {h_2^4 \over \mu^2 r^2} + {2 h_2^2 \over \mu r} - 1} = \\
    = 2 \sqrt{{e_1^2 - 1 \over h_1^2} - {h_1^2 \over \mu^2 r^2} + {2 \over \mu r}} \sqrt{{e_2^2 - 1 \over h_2^2} - {h_2^2 \over \mu^2 r^2} + {2 \over \mu r}} \\
\end{gather*}
We now make the substitutions \eqref{MOaGeom1} and \eqref{MOaGeom2}:

$$
2 \sqrt{{e_1^2 - 1 \over h_1^2} - {h_1^2 \over \mu^2 r^2} + {2 \over \mu r}} \sqrt{{e_2^2 - 1 \over h_2^2} - {h_2^2 \over \mu^2 r^2} + {2 \over \mu r}} =
$$

$$
2 \sqrt{- {2 \over \mu(Q_1+q_1)} - {2 \mu Q_1 q_1 \over \mu^2 r^2 (Q_1+q_1)} + {2 \over \mu r}} \sqrt{- {2 \over \mu(Q_2+q_2)} - {2 \mu Q_2 q_2 \over \mu^2 r^2 (Q_2+q_2)} + {2
    \over \mu r}} =
$$

\begin{equation}\label{}
{4 \over \mu r^2} \sqrt{(Q_1-r)(r-q_1) \over Q_1 + q_1} \sqrt{(Q_2-r)(r-q_2) \over Q_2 + q_2} \\
\end{equation}
For the second and third term of \eqref{eq:vrel_notfinal} we use again the following equality,
\begin{equation}\label{MOaGeom3}
{e^2 - 1 \over h^2} = {- {2 \over \mu (Q+q)}},
\end{equation}
which can be obtained straightforwardly using the substitutions \eqref{MOaGeom1} and \eqref{MOaGeom2} and was already used to simplify the fourth term.

If we substitute the results \eqref{MOaGeom1}-\eqref{MOaGeom3} as well as formula \eqref{eq:cosb1b2} into expression \eqref{eq:vrel_notfinal} we obtain

\begin{multline}\label{eq:vrel}
    v_{12\pm}(r, \varphi) = \sqrt{2 \mu}\ \Bigg( {2 \over r} -\frac{1}{Q_1+q_1}-\frac{1}{Q_2+q_2} \pm {2 \over r^2} \sqrt{\frac{(Q_1-r) (r-q_1)}{Q_1 + q_1}} \sqrt{\frac{(Q_2 - r) (r - q_2)}{Q_2 + q_2}} \\
    - {2 \over r^2} \sqrt{\frac{Q_1 q_1}{Q_1+q_1}} \sqrt{\frac{Q_2 q_2}{Q_2+q_2}} {\pm s_{12}  \sqrt{\cos ^2 \varphi - \cos^2 i_1} \sqrt{\cos^2 \varphi - \cos ^2 i_2}+\cos i_1 \cos i_2 \over \cos ^2 \varphi} \Bigg)^{1 \over 2},
\end{multline}
which, except for $\mu$, is entirely into geometrical terms.

\section{Data}\label{section:Data}

In order to compute the frequency \eqref{eq:CollRate} we need geometrical data of the orbits and the sizes of the objects. The geometrical data, that is, the orbital parameters, are in the TLE of SATTRACK (US Space Force) \cite{spacetrack_web}. As for the sizes, the situation is not that good.

For 6,186 objects (about 1/3 of the total) there is an average cross-section in DISCOS \cite{esa_discos_web} (where it is denoted \texttt{xSectAvg}). For these the radius is computed using

\begin{equation}
R_i = \sqrt{{S_{\text{av},i} \over \pi}},
\label{eq:radius_xsect}
\end{equation}
where $S_{\text{av},i}$ is the average cross-section in m$^2$.

Until August 18, 2014, the Space-Track platform published, alongside each cataloged object, the numerical value of its radar cross-section, expressed in square meters. This data came from measurements taken by the United States Space Surveillance Network, the same network of radars and optical instruments that supports the public catalog of objects in orbit. From that date onward, the numerical value was replaced by a classification in three categories: Small, Medium and Large, whose ranges are (0,0.1) m$^2$, (0.1, 1) m$^2$ and (1,$\infty$) m$^2$. Many of these objects have been identified as objects which are also in the DISCOS database and thus their actual radius is known. In order to identify these objects, an identification number is used, the NORAD${\rm{\_}}$CAT${\rm{\_}}$ID of SATTRACK (US Space Force) \cite{spacetrack_web} and the satno of DISCOS \cite{esa_discos_web}. Not all objects can be assigned a radius using the DISCOS database because when the object is a fragment, it usually doesn't have a radius given by DISCOS. The average and variance of the radii of these known objects are given in the following table:
\medskip

\noindent
\begin{tabular}{|c|c|c|c|c|}
\hline
\multicolumn{5} { | c | }{Non maneuverable objects in DISCOS}\\
\hline
\hline
   Category & Boundaries (m) &$n$ & $\langle R_i \rangle$ (m) & $\langle R_i^2 \rangle$ (m$^2$) \\
\hline
  \hline
  Small &  $\ \ \ \ \ 0 < R_i < 0.18$ & \ \ 557 & 0.1598 & 0.0775 \\
  \hline
  Medium & $\ \ 0.18 < R_i < 0.56$ & 2,052 & 0.4592 & 0.3318 \\
  \hline
  Large & $0.56 < R_i < \infty$ & 3,119 & 1.8949 & 4.3829 \\
  \hline
  Overall & $\ \ \ 0 < R_i < \infty$ & 5,728 & 1.2118 & 2.5130 \\
  \hline
\end{tabular}
\medskip

The total number of objects in the above table is 5,728, that is 458 less than the 6,186 objects mentioned at the beginning of this section. The reason for the discrepancy is that Space-Track didn't assign category to these 458 objects.

Formula \eqref{eq:CollRate} for the collision frequency needs the quantity $(R_1 + R_2)^2$. If none of the radii are known, but it is known that object 1 is in the $i$-th category and object 2 is in the $j$-th category (where $i,\ j \in \{$ small, medium, large $\}$) then $(R_1 + R_2)^2$ is substituted by $\langle(R_i + R_j)^2\rangle = \langle R_i^2 \rangle + 2 \langle R_i \rangle \langle R_j \rangle + \langle R_j^2 \rangle$. If the radius of only one object, say 1, is known, then $(R_1 + R_2)^2$ is substituted by $\langle(R_1 + R_j)^2\rangle = R_1^2 + 2 R_1 \langle R_j \rangle + \langle R_j^2 \rangle$. If the radii of both objects are known, formula \eqref{eq:CollRate} is applied straightforwardly. About $2-3 \%$ of the objects catalogued in SATTRACK do not have an assigned size. To these we have assigned the weighted averages denoted in the table by ``overall''.

The database maintained by Jonathan McDowell (General Catalogue of Artificial Space Objects, GCAT) \cite{McDowell2026GCAT} was used to retrieve the information on the maneuverability of the objects, which decides whether they are included in the present study or not.

\section{Numerical computation}

The collision rate \eqref{eq:CollRate} can only be computed numerically, due to the complexity of its integrand \eqref{eq:density}, \eqref{eq:vrel}. We want to achieve a precision of about 1$\%$, which is more than enough given the uncertainties in the field of space debris. In order to reach this goal, the divergencies of the densities \eqref{eq:density} have to be dealt with at the limits of integration of \eqref{eq:CollRate}, otherwise one would need such a fine grid for the numerical integration that the time required would render the present work impossible to achieve. The treatment of these divergencies is the one given by Kessler in 1981 \cite{Kessler1981}, except that our presentation is more formal and, we hope, more clear. This treatment speeds the computations by a factor of about $10^6$.

It is clear from formulae \eqref{eq:CollRate}, \eqref{eq:density} and \eqref{eq:vrel} that the divergencies come from the pdf's $\rho_1$ or $\rho_2$, not from $\bar v_{12\pm}(r, \varphi)$. To be concrete, of the following six factors

\begin{equation}\label{eq:ProdDiv}
{1 \over \sqrt{(Q_1 - r) (r - q_1) (Q_2 - r) (r - q_2) (\sin^2 I_1 - \sin^2 \varphi) (\sin^2 I_2 - \sin^2 \varphi)}},
\end{equation}
the divergencies are in these three

\begin{equation}\label{eq:ProdDiv}
{1 \over \sqrt{\min(Q_1, Q_2) - r}}, {1 \over \sqrt{r - \max(q_1,q_2)}}, {1 \over \sqrt{\sin^2 \min(I_1,I_2) - \sin^2 \varphi}}.
\end{equation}

To compute the integral we divide the integration region $[\text{max}(q_1,q_2),\ \text{min}(Q_1,Q_2)] \times [0,\ +\min(I_1,I_2)]$ into two regions separated by some $r' \in [\text{max}(q_1,q_2),\ \text{min}(Q_1,Q_2)]$.

\begin{figure}[ht]
\begin{center}
\includegraphics[width=0.75 \textwidth]{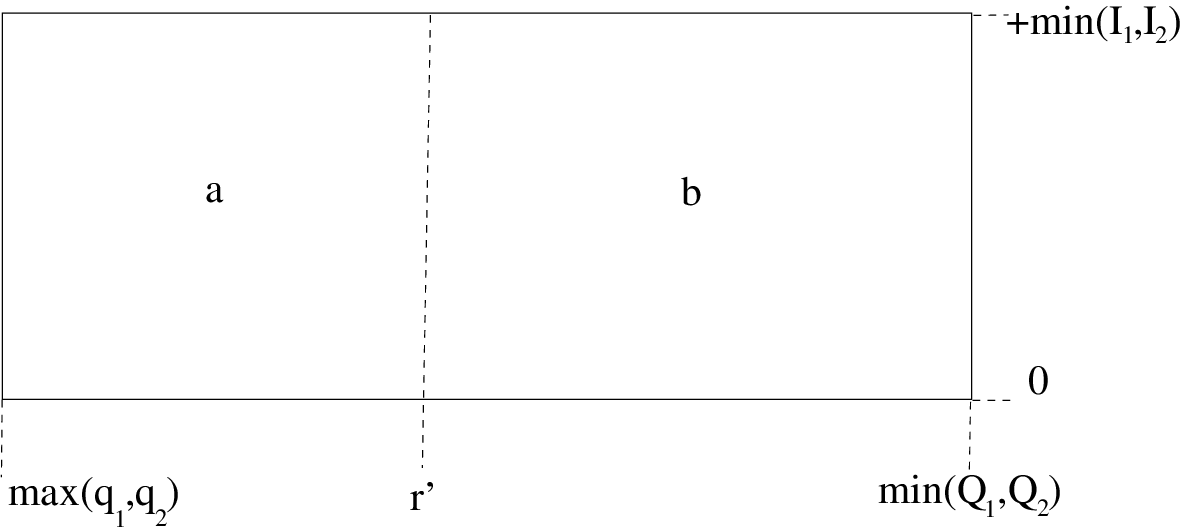}
\caption{\label{fig:Integr} Division of the integration region $[\text{max}(q_1,q_2),\ \text{min}(Q_1,Q_2)] \times [0,\ +\min(I_1,I_2)]$ into two regions separated by some $r' \in [\text{max}(q_1,q_2),\ \text{min}(Q_1,Q_2)]$.}
\end{center}
\end{figure}

The divergencies are due to factors in the integrand which, when isolated, can be integrated analytically. Indeed:

$$
\begin{array}{ccccc}
X_a'(r) & \equiv & {1 \over \sqrt{r-\max(q_1,q_2)}} & \ \ \Rightarrow\ \ & X_a(r) = 2 \sqrt{r-\max(q_1,q_2)} + {\rm{Const.}} \\
X_b'(r) & \equiv & {1 \over \sqrt{\min(Q_1, Q_2)-r}} & \ \ \Rightarrow\ \ & X_b(r) = -2 \sqrt{\min(Q_1, Q_2)-r} + {\rm{Const.}} \\
Y'(\varphi) & \equiv & {\cos \varphi \over \sqrt{\sin ^2 \min(I_1, I_2) - \sin ^2 \varphi}} & \ \ \Rightarrow\ \ &Y(\varphi) = \arctan\left({\sqrt{2} \sin \varphi \over \sqrt{\cos (2\varphi) - \cos (2 \min(I_1, I_2))}}\right) + {\rm{Const.}} \\
\end{array}
$$

\begin{figure}[ht]
\begin{center}
\includegraphics[width=0.75 \textwidth]{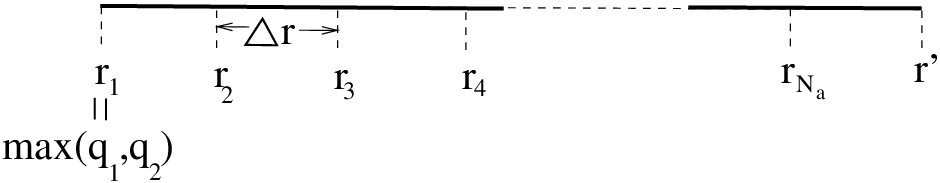}
\caption{\label{fig:intervalos} Division of the $[\text{max}(q_1,q_2),\ r']$ interval into smaller intervals.}
\end{center}
\end{figure}

We divide the integration region $[\text{max}(q_1,q_2),\ r'] \times [0,\ +\min(I_1,I_2)]$ into small rectangles which are the Cartesian product of small intervals such as the ones which are shown in Fig. \ref{fig:intervalos} for the $[\text{max}(q_1,q_2),\ r']$ interval. Similar divisions for the $[r', \text{min}(Q_1,Q_2),]$ and $[0,\ +\min(I_1,I_2)]$ intervals are used.

To avoid the min, max notations, we suppose that $I_1 < I_2,\ q_1 < q_2,\ Q_1 < Q_2$. Then the integral \eqref{eq:CollRate} over the first region, $[\text{max}(q_1,q_2),\ r'] \times [0,\ +\min(I_1,I_2)] = [q_2,\ r'] \times [0,\ I_1]$ is

$$
{2 R^2 \over \pi^4 (q_1 + Q_1) (q_2 + Q_2)} \sum_{j,k=1}^{N_a, N_\varphi}\int_{\varphi_k}^{\varphi_k+\Delta \varphi}d\varphi\ Y'(\varphi)\ \int_{r_j}^{r_j+\Delta r}dr\ X_a'(r)
$$

$$
{\big[ v_{12+}(r, \varphi) + v_{12-}(r, \varphi) \big] \over \sqrt{(Q_1 - r) (r - q_1) (Q_2 - r) (\sin^2 I_2 - \sin^2 \varphi)}} =
$$

$$
{2 R^2 \over \pi^4 (q_1 + Q_1) (q_2 + Q_2)} \sum_{j,k}\int_{\varphi_k}^{\varphi_k+\Delta \varphi} d\varphi {Y'(\varphi)\ \over \sqrt{\sin^2 I_2 - \sin^2 \varphi}}
$$

\begin{equation}\label{}
\int_{r_j}^{r_j+\Delta r}dr {X_a'(r) \over \sqrt{(Q_1 - r) (r - q_1) (Q_2 - r)}} \big[ v_{12+}(r, \varphi) + v_{12-}(r, \varphi) \big].
\end{equation}
At this point we approximate the integrand of every interval by the value that it takes at its center, $\l( r_j+{\Delta r \over 2}, \varphi_k+{\Delta \varphi \over 2} \r)$, except for $X_a'$ or $X_b'$ and $Y'$, which we integrate analytically. This approximation is

$$
{2 R^2 \over \pi^4 (q_1 + Q_1) (q_2 + Q_2)} \sum_{j,k} {Y(\varphi_{k+1}) - Y(\varphi_{k}) \over \sqrt{\sin^2 I_2 - \sin^2 \Big( \varphi_k+{\Delta \varphi \over 2} \Big)}}
$$

$$
{ \l(X_a(r_{j+1})- X_a(r_{j}) \r) \left[ v_{12+}(r_j+{\Delta r \over 2}, \varphi_k+{\Delta \varphi \over 2}) + v_{12-}(r_j+{\Delta r \over 2}, \varphi_k+{\Delta \varphi \over 2}) \right] \over \sqrt{\Big(Q_1 - r_j-{\Delta r \over 2}\Big) \Big(r_j+{\Delta r \over 2} - q_1\Big) \Big(Q_2 - r_j-{\Delta r \over 2}\Big)}} =
$$

\begin{equation}\label{}
C \sum_{j,k} \phi\l( \varphi_k+{\Delta \varphi \over 2} \r) \Re_a\l( r_j+{\Delta r \over 2} \r) f_v\Big(r_j+{\Delta r \over 2}, \varphi_k+{\Delta \varphi \over 2}\Big).
\end{equation}
where
\begin{equation}\label{eq:Constantes0}
\left.
\begin{array}{ccc}
C &\equiv& {\sqrt{2 \mu} R^2 \over \pi^4 (Q_1+q_1) (Q_2+q_2)} \\
\phi(\varphi) &\equiv& {Y(\varphi + {\Delta \varphi \over 2}) - Y(\varphi - {\Delta \varphi \over 2}) \over \sqrt{\sin^2 I_2 - \sin^2 \varphi}} \\
\Re_a(r) &\equiv& { X_a(r+{\Delta r \over 2})- X_a(r-{\Delta r \over 2}) \over \sqrt{(Q_1 - r) (r - q_1) (Q_2 - r)}} \\
\Re_b(r) &\equiv& { X_b(r+{\Delta r \over 2})- X_b(r-{\Delta r \over 2}) \over \sqrt{(Q_2 - r) (r - q_1) (r - q_2)}} \\
f_v(r, \varphi) &\equiv& {v_{12+}(r, \varphi) + v_{12-}(r, \varphi) \over \sqrt{2 \mu}} \\
\end{array}
\right\}.
\end{equation}

The constant $C$ is not the only quantity which takes the same value for all $j,k$. Other constants are:

\begin{equation}\label{eq:Constantes1}
\left.
\begin{array}{cccc}
  Cm & = & \cos (2 \min(I_1, I_2)) \\
  SI_{k} & = & \sin^2 i_k & (k=1,2) \\
  C_3 & = & \frac{1}{Q_1+q_1}+\frac{1}{Q_2+q_2} & \\
  C_4 & = & {2 \over \sqrt{(Q_1+q_1) (Q_2+q_2)}} &  \\
  C_5 & = & 2 \sqrt{\frac{Q_1 q_1}{Q_1+q_1}} \sqrt{\frac{Q_2 q_2}{Q_2+q_2}} & \\
  CI & = & \cos i_1 \cos i_2 & \\
  CI_{k} & = & \cos^2 i_k & (k=1,2) \\
\end{array}
\right\},
\end{equation}
in terms of which the functions $f_v$ and $\phi$ are:

\begin{multline}\label{}
    f_v(r, \varphi) = \sum_{\sigma=-}^{+} \Bigg( {2 \over r} - C_3 + \sigma {C_4 \over r^2} \sqrt{(Q_1 - r) (r - q_1) (Q_2 - r) (r - q_2)} \\
    - {C_5 \over r^2} {\sigma s_{12}(r)  \sqrt{\cos^2 \varphi - CI_{1}} \sqrt{\cos^2 \varphi - CI_{2}} + CI \over \cos ^2 \varphi} \Bigg)^{1 \over 2}
\end{multline}
and

\begin{equation}\label{eq:deffi}
\phi(\varphi) =
{\arctan\left({\sqrt{2} \sin (\varphi + {\Delta \varphi \over 2}) \over \sqrt{\cos (2(\varphi + {\Delta \varphi \over 2})) - Cm}}\right) -
 \arctan\left({\sqrt{2} \sin (\varphi - {\Delta \varphi \over 2}) \over \sqrt{\cos (2(\varphi - {\Delta \varphi \over 2})) - Cm}}\right)
\over \sqrt{SI_2 - \sin^2 \varphi}}
\end{equation}
For completeness we also rewrite $\Re_a$ and $\Re_b$ as

\begin{equation}\label{eq:defRa}
\Re_a(r) = 2 { \sqrt{r+{\Delta r \over 2}-q_2} - \sqrt{r-{\Delta r \over 2}-q_2} \over \sqrt{(Q_1 - r) (r - q_1) (Q_2 - r)}}
\end{equation}
and

\begin{equation}\label{eq:defRb}
\Re_b(r) = 2 { \sqrt{Q_1 - r+{\Delta r \over 2}} - \sqrt{Q_1 -r-{\Delta r \over 2}} \over \sqrt{(Q_2 - r) (r - q_1) (r - q_2)}}.
\end{equation}

The computation of the said constants saves time because they are computed just once and called $N_\phi (N_a + N_b)$ times, as opposed to $N_\phi (N_a + N_b)$ computations and no calls. In order to exploit the same idea further, we define the couple of functions

\begin{equation}\label{}
g_\pm(r) \equiv {2 \over r} - C_3 \pm {C_4 \over r^2} \sqrt{(Q_1 - r) (r - q_1) (Q_2 - r) (r - q_2)}
\end{equation}
and pre-compute

$$
\left\{ \phi\l( \varphi_k+{\Delta \varphi \over 2} \r) \right\}_{k=1}^{N_\varphi},\ \left\{ g_\pm\l( r_j+{\Delta r \over 2} \r) \right\}_{j=1}^{N_a+N_b},\ \left\{ \Re_a\l( r_j+{\Delta r \over 2} \r) \right\}_{j=1}^{N_a}
$$
and
\begin{equation}\label{}
\left\{ \Re_b\l( r_j+{\Delta r \over 2} \r) \right\}_{j=1}^{N_b}.
\end{equation}
In this way we compute $N_\varphi, 2 (N_a + N_b), N_a, N_b$ quantities respectively, which yields a total of $N_\varphi + 3 (N_a + N_b)$ computations and $3 N_\varphi (N_a + N_b)$ calls, as opposed to $3 N_\varphi (N_a + N_b)$ computations and no calls.

Bear in mind that, to lighten the notation, in expressions \eqref{eq:Constantes0}, \eqref{eq:deffi}, \eqref{eq:defRa} and \eqref{eq:defRb} we have supposed that $I_1 < I_2,\ q_1 < q_2,\ Q_1 < Q_2$. When this is not the case, one should substitute $I_1$ by min$(I_1, I_2)$, $I_2$ by max$(I_1, I_2)$, $q_1$ by min$(q_1, q_2)$, $q_2$ by max$(q_1, q_2)$, $Q_1$ by min$(Q_1, Q_2)$ and $Q_2$ by max$(Q_1, Q_2)$.

\section{Results} \label{section:Results}

\subsection{Frequencies of collision} \label{subsection:Frequencies of collision}

The frequency obtained by Kessler \cite{Kessler1978} together with the frequencies that we have obtained are (in units of 1/year) (1976, 0.013), (2016, 0.078), (2021, 0.114), (2026, 0.116). Relevant additional information is contained in the table that follows. A remark about its first line is in order. In Kessler's 1978 work \cite{Kessler1978} evasive maneuvers are not contemplated. Thus all objects are treated as defunct. The number of overlapping couples is not given, because he worked with a sample, as opposed to considering all pairs, as we have.

\medskip
\noindent
\begin{tabular}{|c|c|c|c|c|c|}
\hline
\multicolumn{6} { | c | }{Table 1: Collision frequencies}\\
\hline
  \hline
  Year & Total & Defunct ($\%$) & Defunct pairs & Overlap. def. pairs ($\%$) & Frequency\\
  \hline
  \hline
  1976 & \ 3,866 & \ 3,866 \ \ \ \ \ \ \ \ \ \ \ \ &  &  & 0.013 \\
  \hline
  2016 & 15,532 & 13,769 (88.6 $\%$) & \ 94,785,796 & 39,788,471 (42.0 $\%$) & 0.191 \\
  \hline
  2021 & 21,539 & 17,633 (81.9 $\%$) & 155,452,528 & 58,081,155 (37.4 $\%$) & 0.258 \\
  \hline
  2026 & 30,922 & 16,054 (51.9 $\%$) & 128,857,431 & 42,106,183 (32.7 $\%$) & 0.257 \\
  \hline
\end{tabular}
\medskip

We suppose that the times at which collisions take place are a Poisson process. Then a number of questions may be answered. In particular: if the collision frequency is $1 \over \tau$, what is the probability that at least one collision takes place in the next $t_0$ years? The answer, according to the general theory of Poisson processes \cite{Moran1968} is

\begin{equation}\label{}
1 - \exp -t_0/\tau.
\end{equation}
For example, if $\tau = 1/0.257$ years and $t_0 =$ 5 years, then the probability that a collision takes place in the following 5 years is $1 - \exp {5 \over 1/0.257} \approx 0.723349$. As for the converse question, how long do we have to wait to reach a probability $p_0$ that there is a collision?, the answer is

\begin{equation}\label{}
\tau \ln {1 \over 1 - p_0}.
\end{equation}
For example, if $\tau = 1/0.257$ years and $p_0=0.9$, there is 90$\%$ chance that the next collision takes place in the next $(1/0.257) \ln {1 \over 1 - 0.9}$ years $\approx$ 9 years.

\subsection{Objects most likely to collide}

For the $i$-th object its frequency of collision with any other object is

\begin{equation}\label{}
{1 \over \tau_i} = \sum_{j (j \neq i)} {1 \over \tau_{ij}}.
\end{equation}
If, using the preceding formula, we build a list with the objects which are more likely to suffer a collision, would that be a list of the ``most wanted'' objects to bring down? The answer is no.

The ``most wanted'' objects to bring down are the most dangerous in the sense of debris producing. This danger is, in a loose sense, a product of several factors, the probability of collision being one of them but not the only one. There are two other magnitudes which increase the danger of the objects: the size of the object, because, if broken it will produce more debris, and its altitude, which, like its size, makes it stay longer in orbit.

However, our list of the 46 objects \cite{Estanislao2026} which have the greatest collision probability in May 2026 still shares 17 objects with the list \cite{newspaceeconomy.ca/2025}. The frequencies of collision of our 46 objects are in the range 0.0004-0.001 per year, whereas the range of the objects in the list is 0.000039-0.00023 per year.

\noindent

\section{Comparison with observed frequencies}\label{Comparison with observed frequencies}
%
%

\begin{tabular}{|c|ccc|c|c|}
\hline
\multicolumn{6} { | c | }{Table 2: Confirmed collisions ({\bf{m}} stands for ``maneuverable'' and d for ``defunct'') }\\
\hline
  \hline
  Date & \multicolumn{3} { | c | }{Colliding objects} & Altitude & Fragments \\
  \hline
  \hline
  1991/12/23 \cite{Pardini2014} & Cosmos 1934 (d) & vs. & Cosmos 296 (d) & 980 km & $2$\\
  \hline
  1996/07/24 \cite{Alby1997} & Cerise ({\bf{m}}) & vs. & Ariane stage (d) & 670 km  & 1 \\
  \hline
  2005/01/17 \cite{Cimmino2021} & Thor burner (d) & vs. & Chinese stage (d) & 885 km & $5-7$ \\
  \hline
  2009/02/10 \cite{Wang2010} & Iridium 33 ({\bf{m}}) & vs. & Cosmos 2251 (d) & 776 km & $\approx 2370$ \\
  \hline
  2013/01/22 \cite{Kelso2013} & BLITS ({\bf{m}}) & vs. & Fengyun-1C debris (d) & 830 km & 2 \\
  \hline
  2013/05/23 \cite{Nader2014} & Cubesat NEE-01 (d) & vs. & Tsyklon-3 debris (d) & 645 km & 0 \\
  \hline
  2016/08/23 \cite{Krag2017} & Sentinel 1-A ({\bf{m}}) & vs. & unknown debris (d) & 697 km & $7$ \\
  \hline
  2021/03/18 \cite{Miller2021} & Yunhai 1-02 ({\bf{m}}) & vs. & Russian debris (d)  & 783 km & 37 \\
  \hline
\end{tabular}
\medskip

On 1979/02/06 there was a collision between a maneuverable object (Ayame 1) and a defunct object (the 3rd stage of its own launcher). Like NASA, we don't count it as a fragmentation due to collision, but rather as a fragmentation due to faulty deployment. The ESA, however, lists this fragmentation as a collision \cite{Bongers2025}. The 2013/05/23 collision is a debatable case for two reasons. One is that we have classified it as non-maneuverable, while cubesats have a limited maneuverability because they have attitude control. The other reason is that no fragments were produced. There was however, a collision with a cloud of paint particles released by Tsyklon-3 upper stage which caused the cubesat to spin out of control \cite{Nader2014}.\footnote{Remarkably, the Ecuadorian Civilian Space Agency managed to render this cubesat functional again with the help of a second cubesat \cite{Nader2014}}

\noindent
\begin{tabular}{|c|ccc|c|c|}
\hline
\multicolumn{6} { | c | }{Table 3: Suspected collisions ({\bf{m}} stands for ``maneuverable'' and d for ``defunct'')}\\
\hline
  \hline
  Date & \multicolumn{3} { | c | }{Colliding objects} & Altitude & Fragments \\
  \hline
  \hline
  2020/07/12 \cite{OrbitalDebris2020} & 2018-084C (d) & vs. & ? & \ \ \ 620 km & $74$\\
  \hline
  2023/08/10 \cite{esa_vespa_2023} & VV2 Vega (d) & vs. & ? & \ \ \ 725 km  & ? \\
  \hline
  2026/01/02 \cite{satellitetoday_spainsat_2026} & SpainSat NG II ({\bf{m}}) & vs. & ? & 50,000 km & ? \\
  \hline
  2026/01/30 \cite{jones_luch_2026} & Russian Luch/Olymp (d) & vs. & ? & 43,000 km & ? \\
  \hline
\end{tabular}
\medskip

It is remarkable that perhaps two collisions have taken place at geostationary altitudes in less than a month.

Let us call $\lambda = 0.236$ the average of the frequencies in 2016, 2021 and 2026. If we take the time $t$ to be the 10 year interval from May 2016 to May 2026, then the probability of 0 collisions is $e^{-\lambda t} = 0.094$, the probability of 1 collision is $\lambda t e^{-\lambda t} = 0.223$, the probability of 2 collisions is ${\lambda^2 t^2 \over 2!}e^{-\lambda t} = 0.263$ and the probability of 3 collisions is ${\lambda^3 t^3 \over 3!}e^{-\lambda t} = 0.207$. The tables show that in the said 10 year interval there have been 0 confirmed collisions and 3 suspected collisions between defunct objects, thus the results of our calculations are compatible with the observations.

\section{Conclusions}

We see in formula \eqref{eq:CollRate} that the collision rate $1/\tau_{ij}$ is the product of two factors. The first factor depends only on the size of the colliding objects, and is proportional to $(R_i + R_j)^2$. The second factor depends only on the orbital parameters of the $i$-th and $j$-th bodies. We suppose that the geometry of the drifting objects doesn't change too much in the sense that the distributions of altitude and inclination are roughly constant over the years; in other words we suppose that the average of the second factor has not changed since 1976. If we further suppose that the sizes and the orbital parameters of the defunct bodies are uncorrelated, then we may arrive at a simple approximate formula for the overall collision frequency.



%

If $n$ is the number of drifting objects and $\langle\ \ \ \rangle$ stands for expected value, then

$$
{1 \over \tau} = \sum_{i<j} {1 \over \tau_{ij}} \propto
$$

\begin{equation}\label{}
{n \choose 2} \langle (R_i + R_j)^2 \rangle = {n \choose 2} \langle R_i^2 + R_j^2 + 2 R_i R_j \rangle = n (n-1) \big( \langle R^2 \rangle + \langle R \rangle^2 \big) \approx n^2 \big( \langle R^2 \rangle + \langle R \rangle^2 \big),
\end{equation}
The fraction of pairs which may collide in the years 2016, 2021 and 2026 is a number between 0.32 and 0.42 of the total number of pairs. This fraction doesn't change wildly and thus we take the number of colliding pairs to be roughly proportional to $n^2$. We give the information on $n$, $\langle R^2 \rangle + \langle R \rangle^2$ (in m${}^2$) and the frequency (in 1/years) in the following table. The quantity $\langle R^2 \rangle + \langle R \rangle^2$ for 1976 was computed from Fig. 3 of reference \cite{Kessler1978}, for the other years from the data presented in section \ref{section:Data}.

\medskip
\noindent
\begin{tabular}{|c|c|c|c|}
\hline
\multicolumn{4} { | c | }{Table 4: Data to be fitted by a straight line}\\
\hline
  \hline
  Year & $n$ & $ \langle R^2 \rangle + \langle R \rangle^2 $ & Frequency\\
  \hline
  \hline
  1976 & \ \ 3,866  & 1.42 & 0.013 \\
  \hline
  2016 & 13,769  & 2.58 & 0.191 \\
  \hline
  2021 & 17,633  & 1.97 & 0.258 \\
  \hline
  2026 & 16,054  & 1.37 & 0.257 \\
  \hline
\end{tabular}
\hspace{0.5cm}
\begin{figure}[ht]
\begin{center}
\includegraphics[width=0.9 \textwidth]{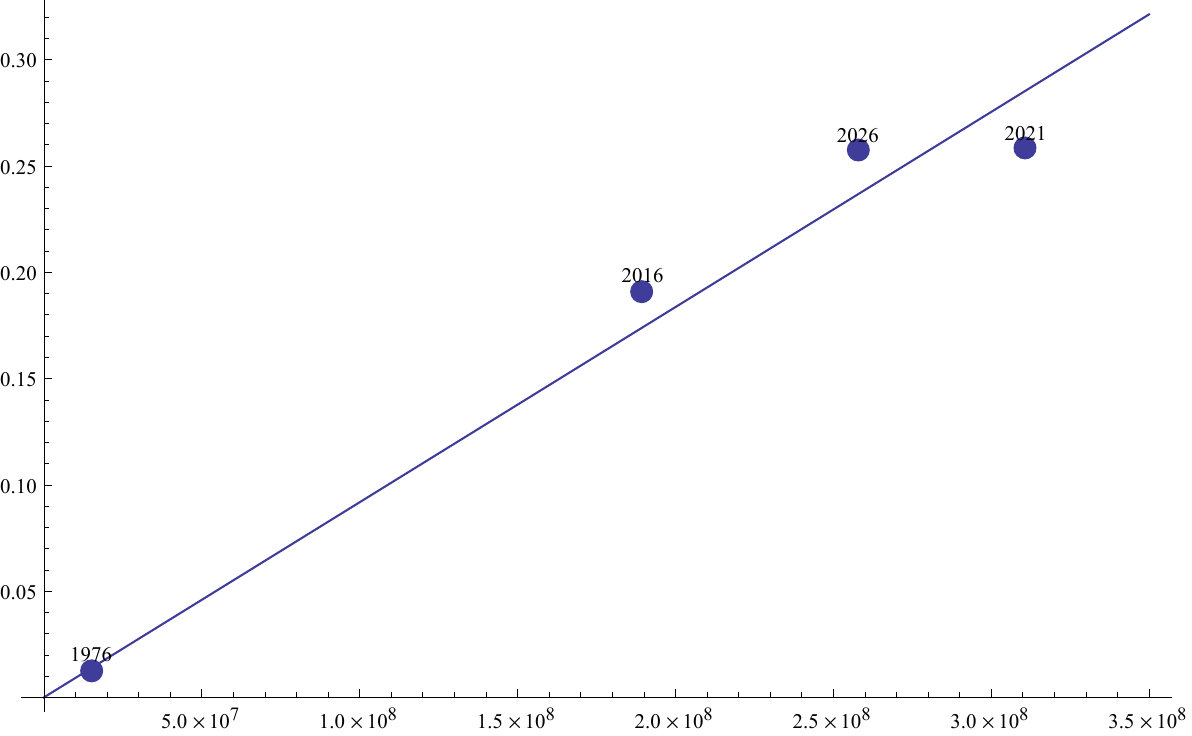}
\caption{\label{fig:fit} Graph of the frequency (in year$^{-1}$) versus $n^2 \big( \langle R^2 \rangle + \langle R \rangle^2 \big)$ (in m${}^2$).}
\end{center}
\end{figure}
\medskip


If we fit a straight line in the plane ($n^2 \big( \langle R^2 \rangle + \langle R \rangle^2 \big)$, frequency) which goes through the origin we obtain

\begin{equation}\label{}
{1 \over \tau} \approx 9.2 \times 10^{-10} n^2 \big( \langle R^2 \rangle + \langle R \rangle^2 \big),
\end{equation}
which we believe might give the frequency in units of 1/year with an error of up to 10$\%$ in the years from 1970 to, say, 2040, provided that the geometry of the drifting objects does not change drastically before 2040. The slope $9.2 \times 10^{-10}$ m$^{-2}$ year$^{-1}$ is the average value of the double integral in formula \eqref{eq:CollRate} times $2 \pi^2$.

The collision frequency has increased from 2016 to 2021, but (within error) stayed the same in the last 5 years. Thus the frequencies obtained for the last 10 year period are inconclusive with regard to the question of whether the Kessler cascading is taking place or not. Hopefully this computation, with better or new data, will help to answer this question.

It is remarkable that the computation of the collision frequency among orbiting objects of known orbital parameters and sizes had not been, as far as we know, repeated since Kessler's 1981 paper \cite{Kessler1981}. We hope that this article, where this computation is spelled out at length, encourages researchers to compute this quantity, not only to check whether Kessler's cascading is taking place or not, but for other purposes also.

\bigskip

\vspace{1in}
\newpage

\section{Appendix I: Keplerian velocity}

%

%
%
%

A formula for the velocity in a Keplerian elliptical orbit in polar coordinates with origin at the occupied focus and positive direction along the eccentricity vector $\vec e$ is

$$
\dot {\vec r}(t) = \dot r \vec u_r + r \dot \theta \vec u_\theta = \dot \theta \left( {d r \over d \theta} \vec u_r + r \vec u_\theta \right) =
h \left( {1 \over r^2} {d r \over d \theta} \vec u_r + {1 \over r} \vec u_\theta \right) =
$$

\begin{equation}\label{}
h \left( - {d (1/r) \over d \theta} \vec u_r + {1 \over r} \vec u_\theta \right).
\end{equation}
We use the substitutions

\begin{equation}\label{}
\left.
\begin{array}{c}
{1 \over r} = {\mu \over h^2} (1 + e \cos \theta) \\
{d (1/r) \over d \theta} = - {\mu \over h^2} e \sen \theta
\end{array}
\right \}
\end{equation}
and obtain

\begin{equation}\label{eq:VelKepleriana1}
\dot {\vec r}(t) = {\mu \over h} \left( e \sen \theta\ \vec u_r + (1 + e \cos \theta) \vec u_\theta \right) = {\mu \over h} e \left( \sen \theta\ \vec u_r + \cos \theta\ \vec u_\theta \right) + {\mu \over h} \vec u_\theta.
\end{equation}

\begin{figure}[ht]
  \begin{center}\label{fig:VelKeplerianas3.eps}
  \includegraphics[width=0.6 \textwidth]{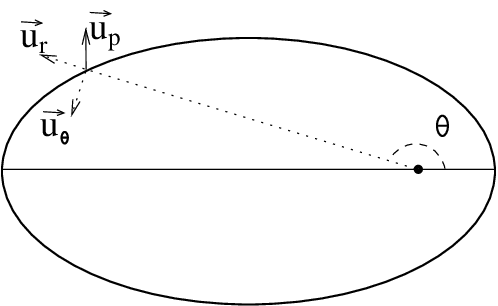} \caption{$\vec u_\theta$ is the unit vector perpendicular to the radius vector. It is not the unit vector tangent to the trajectory. $\vec u_p$ is the unit vector perpendicular to the major axis, such that $\vec u_e \wedge \vec u_p = \vec u_h$.}
  \end{center}
\end{figure}

The vector $\sen \theta\ \vec u_r + \cos \theta\ \vec u_\theta $ is a unit vector perpendicular to the vector $\vec u_e$ (because $\vec u_e \cdot (\sen \theta\ \vec u_r + \cos \theta\ \vec u_\theta) = \sen \theta \cos \theta - \cos \theta \sen \theta = 0$) and such that $\vec u_e \wedge \left( \sen \theta\ \vec u_r + \cos \theta\ \vec u_\theta \right) = \vec u_h$, therefore it is the vector $\vec u_p$ of the perifocal frame, $\vec u_p = \left( \sen \theta\ \vec u_r + \cos \theta\ \vec u_\theta \right)$. Then

\begin{equation}\label{eq:VelKeplerianaApp}
\dot {\vec r}(t) = {\mu \over h} \left( e \vec u_p + \vec u_\theta \right) = \sqrt{\mu} \sqrt{q+Q \over 2 qQ} \left( e \vec u_p + \vec u_\theta \right).
\end{equation}

\section{Appendix II: Circular orbits}

When both orbits are circular $h_1 = h_2 = \sqrt{\mu r},\ e_1 = e_2 = 0$, the velocities have no vertical components and expression \eqref{eq:vh12} simplifies to
\begin{equation}\label{}
v_{12} = \mu \left( {1 \over h_1^2} + {1 \over h_2^2} - 2 {1 \over h_2} {1 \over h_1} \cos (b_2 - b_1) \right)^{1 \over 2} =
{\sqrt{2} \mu \over h} \left( 1 - \cos (b_2 - b_1) \right)^{1 \over 2}.
\end{equation}

\begin{figure}[ht]
\begin{center}
\includegraphics[width=0.45 \textwidth]{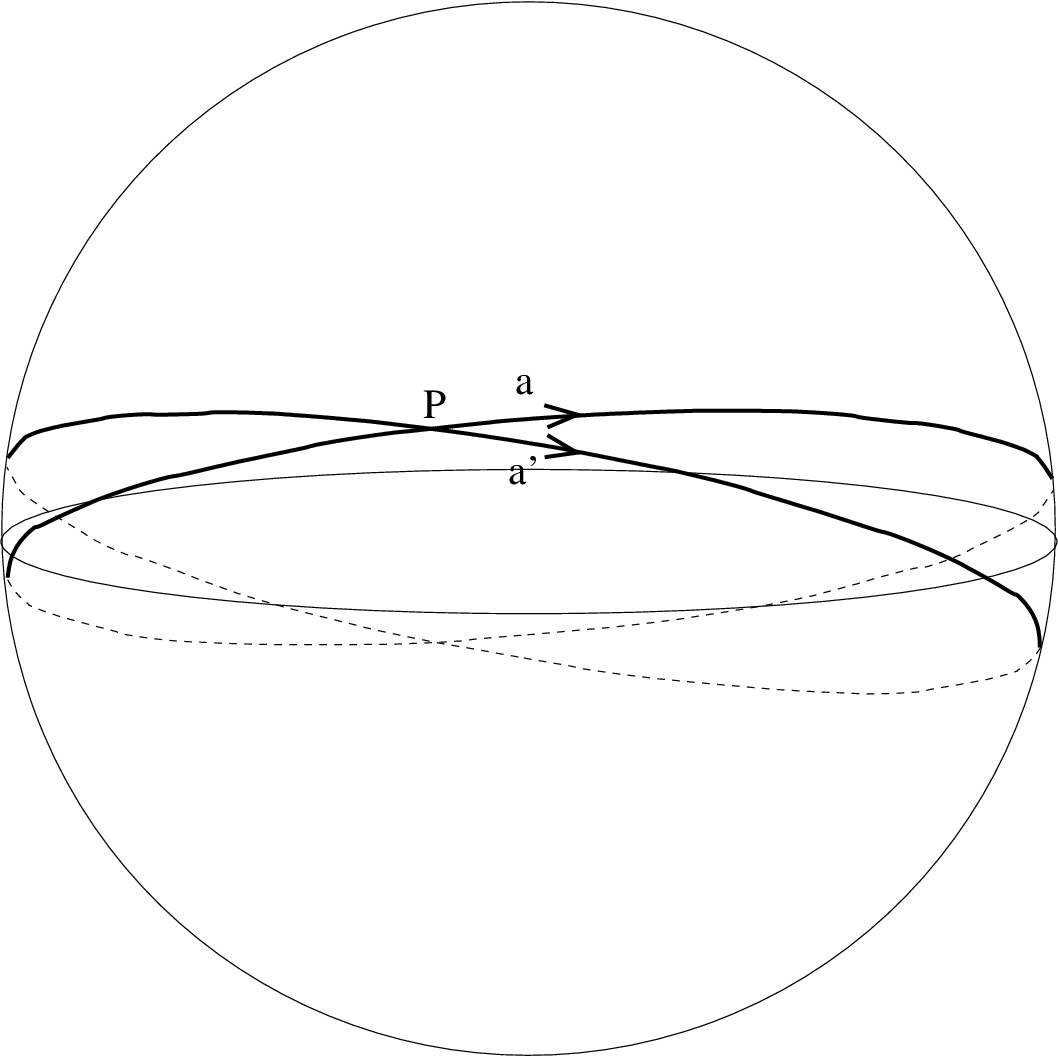}\hspace{0.07 \textwidth}\includegraphics[width=0.37 \textwidth]{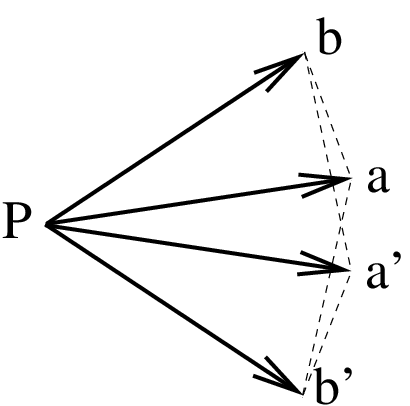}
\caption{\label{fig:VelCirc} In the left figure the two circular orbits with given $i$ which go through a point $P$ are shown. At $P$ the body in orbit $a$ is yet to reach the point of maximum latitude, while the body in orbit $a'$ is already past it. In the right figure the velocity vectors of orbits $a$ and $a'$ are shown, as well as the two possible velocity vectors of an orbit with a different inclination. It is clear from the right figure that while there are four possible relative velocities, there are only two possible relative speeds.}
\end{center}
\end{figure}
In order to make the above formula ready for use we proceed as in section \ref{Relative speed}:

$$
v_{12\pm}(r, \varphi) =
$$

\begin{equation}\label{eq:v12circ}
\sqrt{2 \mu \over r}\ \Bigg( 1 - {1 \over \cos^2 \varphi} \Big( \pm  \sqrt{\cos ^2 \varphi - \cos^2 i_1} \sqrt{\cos^2 \varphi - \cos ^2 i_2} + \cos i_1 \cos i_2 \Big) \Bigg)^{1 \over 2},
\end{equation}
where the two signs correspond to the two possible speeds shown in Fig. \ref{fig:VelCirc}. If, furthermore, $i_1 = i_2 = i$, or if $i_1 = i$ and $i_2 = \pi-i$, then it must be $\cos^2 \varphi = \cos^2 i_1 = \cos^2 i_2$ and

\begin{equation}\label{eq:v12circ1}
v_{12\pm}(r) = 0\ \ \ \ {\rm{or}}\ \ \ \ v_{12\pm}(r) = 2 \sqrt{\mu \over r},
\end{equation}
respectively.

Suppose now that only one of the orbits is circular, say orbit 2.
Since $e_2 = 0$, formula \eqref{eq:vh12} becomes

\begin{equation}\label{}
v_{12} =
\mu \left( {1 + e_1^2 \over h_1^2} + {1 \over h_2^2} + 2 {e_1 \cos \theta_1 \over h_1^2} - 2 {1 \over h_2} {1 + e_1 \cos \theta_1 \over h_1} \cos (b_2 - b_1) \right)^{1 \over 2}.
\end{equation}
Note that the term proportional to $e_1 e_2 \sin \theta_1 \sin \theta_2$, which gave rise to a double sign, has disappeared. This is because $\vec v_2$ is horizontal and $\vec v_1$ takes two values whose vertical components differ only in sign. Therefore there is only one value for the vertical relative speed. As for the horizontal relative speed, it is like the case of two circular velocities, for which we have just seen that there are two possibilities shown in result \eqref{eq:v12circ}. Thus,

\begin{multline}\label{}
    v_{12\pm}(r, \varphi) = \sqrt{2 \mu}\ \Bigg( {1 \over r} -\frac{1}{Q_1+q_1}  \\
    - \sqrt{2 \over r^3} \sqrt{\frac{Q_1 q_1}{Q_1+q_1}} {\Big( \pm \sqrt{\cos ^2 \varphi - \cos^2 i_1} \sqrt{\cos^2 \varphi - \cos ^2 i_2}+\cos i_1 \cos i_2 \Big) \over \cos ^2 \varphi} \Bigg)^{1 \over 2}.
\end{multline}
\vspace{1in}

\bibliography{refer12}
\bibliographystyle{unsrt}

\end{document}